\documentclass[aps,preprint,showpacs,preprintnumbers,amsmath,amssymb]{revtex4}
\usepackage[dvips]{graphicx}
\usepackage{float}
\begin{document}

\title{Reply to the "Comment on Fermion production in a magnetic field in a de Sitter Universe"}
\author{Cosmin Crucean \thanks{E-mail:~~crucean@physics.uvt.ro}
and Mihaela-Andreea B\u aloi \thanks{E-mail:~~mihaela.baloi88@e-uvt.ro}\\
{\small \it West University of Timi\c soara,}\\
{\small \it V. Parvan Avenue 4 RO-300223 Timi\c soara,  Romania}}

\begin{abstract}
In this paper we study the problem of gauge invariance of the first order transition amplitudes in de Sitter QED in Coulomb gauge. We consider the gauge transformations which preserve the Coulomb gauge, that contain the gradient of the gauge function. The final results prove that the first order transition amplitudes do not change at a gradient transformation of the vector potential because the only allowed transformation is $\Lambda=0$. Our results suggest that the remarks made in the comment by Nicolaevici and Farkas {\em Phys. Rev. D} \textbf{95}, are not directly applicable to the results in our paper since their proposed gauge transformations do not preserve the Coulomb gauge.
\end{abstract}

\pacs{04.62.+v}

\maketitle

\section{Introduction}

The problem of gauge invariance of the transition amplitudes in Minkowski QED was the subject of an intense debate some decades ago \cite{2,3,4,5,6,8,9,10,19,20,21,23,24}. While some of the authors argue that the surface terms that appear in the variation of the amplitude in the first order, at a gauge transformation of the potential $A^{\mu}$, are vanishing due to the fact that the fields vanish if we impose boundary conditions \cite{6,10,19,20,21,22}, others prove that the problem of gauge invariance of the amplitudes must be discussed in relation with the renormalization of the Minkowski QED \cite{2,3,4}, since the renormalization constants also depend on the chosen gauge. In quantum field theory on curved spacetime \cite{7} the problem of gauge invariance was not studied in detail, and the existing results do not allow us to reach definitive conclusions regarding the gauge dependence of the amplitudes. The perturbative QED on the de Sitter spacetime was constructed in \cite{14} where it is shown that that a mandatory condition for quantifying the whole theory is to choose and fix the Coulomb gauge.
Moreover, this seems to be a specific feature of the de Sitter geometry as long as the massless limit of the free Proca field on this background \cite{17}, gives just the free Maxwell field in Coulomb gauge.

Another result obtained recently \cite{15}, shows that there are indications that the transition amplitudes in the first order of perturbation theory are gauge dependent in de Sitter spacetime. The comment \cite{16} to our paper \cite{1} is a continuation of the result obtained in \cite{15}, and the cause of the gauge dependence of the amplitudes was indicated to be contained in the temporal part of the Dirac current which loses it's oscillatory behaviour in the infinite future. In this Reply we want to reconsider the problem of gauge invariance of transition amplitudes in de Sitter QED. This is done by discussing the situation from our paper \cite{1}, where we use Coulomb gauge and by making some observations about the results obtained in the comment \cite{16}.

The paper is organized as follows: in the second section we discuss the problem of the transition amplitudes when the Coulomb gauge is used. In the third section the problem of gauge invariance is discussed in relation with the vector potential decomposition in perpendicular and parallel components. In section four we discuss the problem of gauge invariance in relation with the renormalization of the theory and our conclusions are presented in section five.

\section{Amplitudes in Coulomb gauge}

The line element \cite{0} which describes the de Sitter universe is:
\begin{equation}\label{metr}
ds^{2}=dt^{2}-e^{2H t}d\vec{x}\,^{2}=\frac{1}{(H\eta)^2}(d\eta^{2}-d\vec{x}\,^{2})
\end{equation}
where $H$ is the expansion factor and $H>0$, while $\eta=-\frac{1}{H}\,e^{-H t} $ is the conformal time.

It is known that Minkowski QED is a gauge invariant theory in the sense that any transformations of the type:
\begin{eqnarray}\label{tr}
&&A_{\mu}\rightarrow A_{\mu}+\partial_{\mu}\Lambda\nonumber\\
&&\psi\rightarrow e^{ie\Lambda}\psi
\end{eqnarray}
leaves invariant the field equations, where $\Lambda$ is the gauge function which is a scalar function dependent both on time and spatial coordinates \cite{18,19,21,22,23}. The second transformation from (\ref{tr}) refers to the matter fields and this second transformation must be accompanied by the transformation of the potential in the case when the electromagnetic field is coupled with a matter field for leaving unchanged the field equations. We will assume the usual boundary conditions in space and time, such that the gauge function vanishes at infinity \cite{20,22}:
\begin{equation}
\Lambda\rightarrow0,
\end{equation}
for $ t\rightarrow\infty, x\rightarrow\infty$.

Before proceeding in our analysis let us make a few observations related to the Coulomb gauge. It true that in our paper about the fermion production in the field of the magnetic dipole \cite{1}, we omit to mention that we use Coulomb gauge. The reasons for using Coulomb gauge will be detailed in what follows and an extended discussion can also be found in our previous papers \cite{13,14,26}. The significance of Coulomb gauge in de Sitter geometry is very important if we take into account the conformal invariance of the Maxwell equations. Then it seems that the Coulomb gauge is the only gauge which opens the way to conformally relate the whole theory of Maxwell field written in the chart with conformal time $\{\eta,\vec x\}$, to the usual electrodynamics from Minkowski spacetime. The Lorentz condition is conformally invariant only in Coulomb gauge in de Sitter geometry, and this allows us to obtain the solutions of the free Maxwell equations in the helicity-momentum basis as in flat space theory \cite{26}. The canonical quantization of the free Maxwell field can be performed in Coulomb gauge where the Lorentz condition becomes conformally invariant \cite{26} ,as we point out above. This means that it is useful to maintain this gauge for constructing the theory of interacting fields \cite{14}. However there is a lot of work to be done if we want to speak about measurable quantities in this geometry, but an important result is that we recover in the limit of zero expansion factor the results from flat space QED. This means that in the limit of zero expansion factor the transition amplitudes and probabilities computed in de Sitter QED in Coulomb gauge, reduce to those from Minkowski theory.
How the theory of free Maxwell field look in other gauges in de Sitter geometry it is not known to the best of our knowledge and therefore further studies must be done for fully understand the theory of Maxwell field in this geometry. For the above mentioned reasons we restrict ourselves to use the Coulomb gauge for studding the perturbative QED in de Sitter geometry.

All the details related to the construction of perturbative QED in de Sitter geometry using the Coulomb gauge can be found in \cite{14}, and we remind here only the main steps.
The construction of the de Sitter QED in Coulomb gauge starts with the Lagrangian theory that gives the field equations and the principal conserved observables of the interacting fields \cite{14}. Then the equal time commutators and anticommutators are postulated and the equation of the time-dependent evolution operator are derived, obtaining the perturbation series of the scattering operator in terms of free fields. This is generated by the interaction Hamiltonian which does not depend on the Coulomb potential. So the Coulomb gauge allows a natural quantization separating the Coulomb potential \cite{14}. Finally the asymptotic fields are defined and we obtain the $in-out$ amplitudes by using the reduction formalism and the scattering operator \cite{14}.

Thus, one first chooses a gauge for solving the free Maxwell equations, then the quantization can be done. The theory of fields interactions is also constructed by choosing a gauge since we have to find the solutions for the interacting fields equations, which in de Sitter geometry are strongly dependent on gauge \cite{14}. Then the perturbation theory can be constructed and finally we can obtain the expressions for the transition amplitudes in any order. It is then clear that once a gauge is fixed and the quantization procedure is done, one could not do a gauge transformation in the transition amplitudes for passing to another gauge. In these circumstances any gauge transformations related to the electromagnetic potentials that are allowed are just those that preserve the chosen gauge in which the theory of interacting fields was constructed and the quantization was done. The authors of the comment \cite{16} to our paper seem to miss this important observation, since the only allowed gauge transformations are those that preserve the Coulomb gauge and their discussion should consider these transformations only, for commenting on our paper \cite{1}. Instead at the beginning of the comment the authors leave the impression that they work in Coulomb gauge taking into account that they consider $A_{i}\neq0\,,A_{0}=0$ and then they choose to made the discussion in another gauge such that the amplitude is defined in general with $A_{\mu}$. This comment should construct the theory of free electromagnetic field in other gauge, in de Sitter geometry, then make the quantization and finally construct the QED in this new gauge. Therefore a comment to our papers \cite{1,13,14} should prove the following: first take another gauge and solve the Maxwell equations and then construct the perturbative QED in this new gauge. Then in this new gauge take a potential which give the same dipolar magnetic field (as the one used in our paper \cite{1}), and compute the first order amplitude corresponding to the fermion pair production and finally compare the results from the this new gauge with our results obtained in Coulomb gauge. To conclude, the comment \cite{16} did not prove that the amplitudes from our paper \cite{1}, are gauge dependent. In what follows we will clarify what it means to do a gauge transformation in our case \cite{1}.

The amplitude of pair production in an external magnetic field from our paper \cite{1}, will be further considered.
In \cite{1}, the vector potential that describe the magnetic field produced by a dipole was taken as
\begin{equation}\label{am}
\vec{A}=\frac{\vec{\mathcal{M}}\times\vec{x}}{|\vec{x}|^3}\,e^{-H t},\,\,A^{0}=0,
\end{equation}
where $\vec{\mathcal{M}}$ is the magnetic dipole moment. We also observe that $\nabla\vec{A}=0,\,A_{0}=0$. The first order transition amplitude corresponding to pair creation in external field, assuming the minimal coupling, is \cite{1}:
\begin{equation}\label{ampl1}
\mathcal{A}_{e^-e^+}=-ie \int d^{4}x
\left[-g(x)\right]^{1/2}\bar U_{\vec{p},\,\lambda}(x)\vec{\gamma}\cdot\vec{A}(x) V_{\vec{p}\,\,',\,\lambda'}(x),
\end{equation}
where $U_{\vec{p},\,\lambda}(x),V_{\vec{p}\,\,',\,\lambda'}(x)$ are the solutions of the Dirac equation in momentum basis in de Sitter geometry \cite{12}.

Our analysis in de Sitter case is done preserving the Coulomb gauge, and the same is true for our paper \cite{1}. Let us denote by $\nabla_{\alpha}$ the covariant derivative, with $\alpha=0,1,2,3$. The condition $\partial_{i}A^{i}=0$ from Minkowski space is replaced by the vanishing of the covariant derivative in de Sitter case $\nabla_{i}(\sqrt{-g}A^{i})=0$ \cite{14,26}, but it is sufficient to apply the covariant derivative only on $A^{i}$ since $\sqrt{-g}$ depends only on time and we obtain:
\begin{equation}
\nabla_{i}A^{i}=0=\partial_{i}A^{i}+\Gamma^{i}_{i\alpha}A^{\alpha}=\partial_{i}A^{i}+\Gamma^{i}_{ij}A^{j}+\Gamma^{i}_{i0}A^{0}
\end{equation}
Considering the de Sitter line element (\ref{metr}), with $g_{00}=1,\,g_{ij}=-\delta_{ij}e^{2H t}$, we obtain that $\Gamma^{i}_{ij}=0,\,\Gamma^{i}_{i0}=H$ and the above equation becomes:
\begin{equation}
\nabla_{i}A^{i}=\partial_{i}A^{i}+H A^{0}=0,
\end{equation}
since $A^{0}=0$ this implies $\partial_{i}A^{i}=0$.

It is known that further gauge transformation that preserve the Coulomb gauge condition can be made,
and these gauge transformations which contain the components of $A^{\mu}$ can be written as:
\begin{eqnarray}\label{at}
A^{i}\rightarrow A^{i}+\partial^{i}\Lambda\,\nonumber\\
A^{0}\rightarrow A^{0}+\partial^{0}\Lambda,
\end{eqnarray}
since $A^{0}=0$ we observe that the condition $\partial^{t}\Lambda=0$ is mandatory (this conclusion can be reached following similar arguments like in the Minkowski theory, see \cite{9}).

Let us apply the covariant derivative to the potential transformation given in equation (\ref{at}):
\begin{equation}
\nabla_{i}A^{i}\rightarrow\nabla_{i}A^{i}+\nabla_{i}(\partial^{i}\Lambda)
\end{equation}
and compute $\nabla_{i}(\partial^{i}\Lambda)$ to obtain:
\begin{equation}
\nabla_{i}(\partial^{i}\Lambda)=\partial_{i}\partial^{i}\Lambda+\Gamma^{i}_{ij}\partial^{j}\Lambda+\Gamma^{i}_{i0}\partial^{0}\Lambda=\partial_{i}\partial^{i}\Lambda
+H \partial^{t}\Lambda.
\end{equation}
Since we impose the condition $\nabla_{i}A^{i}=0$, we observe that we must take $\nabla_{i}(\partial^{i}\Lambda)=0$ for preserving the Coulomb gauge and finally obtain the equation for $\Lambda$ in de Sitter geometry:
\begin{equation}\label{la}
\partial_{i}^{2}\Lambda-
H e^{2H t}\partial_{t}\Lambda=0.
\end{equation}
A similar situation is encountered in Minkowski QED, where for preserving the Coulomb gauge the condition $\partial_{i}^2\Lambda=0$ is imposed \cite{20,22,23,25}.
In equation (\ref{la}), we observe that $\Lambda$ is a time independent function, i.e. $\partial_{t}\Lambda=0$, as shown above and the first term of the equation reproduce the situation from flat space case, giving for the gauge function an equation of the Laplace type \cite{20,22,23,25}:
\begin{equation}\label{ss}
\triangle\Lambda=0.
\end{equation}

From the analysis above we observe that in the Coulomb gauge the gauge function $\Lambda$ depends only on spatial coordinates. But the above equation (\ref{ss}) has a nice property: its solutions are unique \cite{23}. In other words if we can find a solution to Laplace equation
which satisfies the boundary conditions then it is clear that this is the only solution. The physical criterion that the gauge function must accomplish, is for it to vanish when the spatial distances and the time become infinite, where the Dirac fields and the potential used in our calculations also vanish \cite{19,20,22,23}. In these circumstances the unique solution that accomplishes these criteria is \cite{18,22,23}:
\begin{equation}\label{z}
\Lambda=0
\end{equation}
and no gauge arbitrary remains in this case. So our transition amplitudes of fermion production in the magnetic field on de Sitter spacetime obtained in \cite{1}, do not change if we add to the potential the gradient of the gauge function as long as we impose to remain in the Coulomb gauge. To the best of our knowledge how physics looks in de Sitter geometry if we chose other gauges is not studied in literature. Instead it seems that for the moment the Coulomb gauge is the only gauge in which one could construct the theory of free electromagnetic field imposing then the canonical quantization and further making a coherent perturbative QED \cite{14}.

An interesting observation about our vector potential is that we know it's divergence and curl and in addition we know that it vanishes when the spatial distances become infinite. Then a vector field which vanishes at infinity is completely specified once its divergence and its curl are given ($\nabla\vec{A}=0\,\,,\nabla\times\vec{A}=\vec{B}$) \cite{22}. There are no additional terms due to the transformation (\ref{at}) in our amplitude of fermion production in magnetic field, and this is the result of the use of Coulomb gauge and of the use of boundary conditions in space and time. So we prove that any gradient transformation leaves the transition amplitude invariant in Coulomb gauge. Or in other words the only gauge transformation allowed which preserve the Coulomb gauge is $\Lambda=0$, as in the Minkowski QED \cite{23}.

\section{Gauge invariance and vector potential decomposition}
Another way to tell the above story is as follows. Consider that the vector potential is decomposed in longitudinal and transversal parts \cite{27}:
\begin{equation}
\vec{\mathcal{A}}=\vec{\mathcal{A}}_{\perp}+\vec{\mathcal{A}}_{\parallel}
\end{equation}
such that \cite{27}, $\nabla\vec{\mathcal{A}}_{\perp}=0\,\,,\nabla\times\vec{\mathcal{A}}_{\parallel}=0$, which is known since the magnetic field is purely transversal $\vec{B}_{\parallel}=0$. Then the gauge transformation:
\begin{equation}
\vec{\mathcal{A}}\rightarrow\vec{\mathcal{A}}+\nabla\Lambda
\end{equation}
will give the transformation rules for the longitudinal and transversal components \cite{27}:
\begin{eqnarray}
&&\vec{\mathcal{A}}_{\perp}\rightarrow\vec{\mathcal{A}}_{\perp},\nonumber\\
&&\vec{\mathcal{A}}_{\parallel}\rightarrow\vec{\mathcal{A}}_{\parallel}+\nabla\Lambda.
\end{eqnarray}
We observe from the above equation that the transversal component of the potential vector is gauge invariant \cite{27} and only the longitudinal component transforms. In our calculations we use only the transversal component $\vec{A}=\vec{\mathcal{A}}_{\perp}$, as given in equation (\ref{am}). Moreover in the Coulomb gauge $\vec{A}_{\parallel}=0$ \cite{27}, and only the transversal components are not vanishing.

Finally, we conclude that working in the Coulomb gauge, only with the transversal components of the potential vector one can obtain results which are gauge independent in de Sitter QED. For studying the theory of free electromagnetic field and perturbative QED and then compute the amplitudes one needs to fix a gauge.

\section{Gauge invariance and renormalization}

Since we know from flat space QED that for computing observable quantities one needs to do the renormalization of the theory, which depends on gauge \cite{2,3,4,5,9}, the same observation could be also valid in de Sitter QED. Then the discussion of gauge invariance in de Sitter QED is not at all an easy task since a complete proof of the gauge independence/dependence of the amplitudes/probabilities could depend on the renormalization of the theory, which is an issue not clear at this moment in this geometry. It is known from flat space QED that the unrenormalized $S$-matrix elements obtained from Feynman diagrams
always appear in physical-scattering amplitudes  multiplied by the renormalization constants $Z_2$ and $Z_3$. As we know $Z_3$ is gauge invariant but $Z_2$ is a gauge dependent quantity \cite{2,3,4,5}. This means that the unrenormalized transition amplitudes, could be, gauge dependent in order to secure the gauge invariance of the product between the renormalization constants and the unrenormalized matrix elements \cite{2,3,4,5}. So a proof of the gauge invariance could be carried out on the renormalized amplitudes/probabilities in de Sitter QED. For that we must study the Maxwell and Dirac propagators including their radiative corrections. The above program must be completed, and only then a definite conclusion could be addressed properly about the gauge dependence of the amplitudes in de Sitter QED.

\section{Conclusions}

The final conclusion is that the comment \cite{16}, needs to be considered with care, and the authors do not present valid arguments in what regards the gauge dependence of the amplitude from our paper \cite{1}. In our paper we work in Coulomb gauge in which the only allowed gauge transformations are $\Lambda =0$. In quantum field theory the standard procedure is to establish the gauge first, and only afterwards perform the quantization. Once a gauge is fixed and the quantization procedure is done, one could not do a gauge transformation in the transition amplitudes which alter the chosen gauge.

Another important observation is that the analysis of amplitudes variation in other gauges and with given external electromagnetic fields must be done in de Sitter geometry in order to understand the problem of gauge invariance, but there are no concrete results at the present time in the literature.

Since the authors of the comment do not say or prove how the analytical results or physical interpretation of our results will be modified by their result, we disagree with the implication of the comment \cite{16} that the analysis in our paper \cite{1} and our previous papers \cite{13,14}, are physically incorrect or there are ambiguities about the quantities that were computed.

\textbf{Acknowledgements}

We would like to thank to Professor L. H. Ford for his observations that help us to improve our paper. We would also like to thank Professor Ion Cot\u aescu, Dr. Victor Ambru\c s and Dr. Sporea Ciprian for interesting discussions and for reading our paper.

\end{document}